# Quantum Point Contact Parameter Extraction of Carbon-based Resistive Memory using Hybrid Genetic Algorithm

Ee Wah Lim

**ABSTRACT**

Resistive switching phenomenon in carbon film is associated with formation and annihilation of low resistance $sp^2$ nanochannels within the amorphous $sp^3$ matrix. The thinnest point of these graphitic nanochannels behaves like quantum wire (QW) and limits current flow. Transport mechanism at these bottlenecks can be described within the framework of quantum point contact (QPC) model. The model applies mesoscopic Landauer formalism and correlates device resistance state with the density of the nanochannels as well as lateral area of its constriction. However, QPC model parameter extraction from macroscopic I-V characteristic is not feasible due to multiple nonlinear and closely coupled parameters, *e.g.* barrier height ($\Phi$), barrier curvature ($\alpha$) and voltage drop ratio ($\beta$). In this work, a hybrid genetic algorithm (GA) based parameter extraction flow is proposed and is applied to extract parameters from experimental result of carbon-based resistive memories. In this proposed flow, the number of single subband channels for low resistance state ($N_{LRS}$) and barrier curvature parameter at high resistance state ($\alpha_{HRS}$) were first derived by exploring boundary conditions at low voltage. The four remaining parameters ($\Phi_{HRS}$, $\Phi_{LRS}$, $\alpha_{LRS}$, and $\beta$) were then extracted from experimental I-V characteristic by leveraging hybrid GA. Mean absolute percentage error (MAPE) of fitted QPC models is only 3.7% and the extracted parameters are within reasonable range. Hereby, we regard the QPC model as a viable model to describe conduction mechanism of graphitic nanochannels in carbon-based resistive memories.

**Keywords**
QPC model; Carbon-based Memory; RRAM; Genetic Algorithm; Parameter Extraction

## 1. Introduction

The scaling of conventional charge-based non-volatile memory (NVM) is approaching its inherent physical scaling limitation [1]. As a consequence, resistive random access memory (RRAM) is said to be one of the most promising emerging memory technologies due to its simple structure, high performance and compatibility with the conventional semiconductor process [2]. RRAM is essentially a bi-stable metal-insulator-metal (MIM) device and is often referred to as memristor [3]. Resistive switching phenomena have been reported in a wide range of materials, *e.g.* transition metal oxide (TMO), organic materials, perovskite oxide and carbon-based materials. Among these materials, carbon-based films have unique advantages such as controllable preparation process, low fabrication cost and high temperature retention of 250°C, which makes it attractive for some harsh applications such as automotive [4]. Due to the impractical forming process of the conductive carbon film [4], typical carbon-based resistive memories are deposited in an insulating form, such as graphene oxide (GO) [5,6], amorphous carbon (a-C) [7], hydrogenated a-C (a-C:H) [8,9] and diamond-like carbon (DLC) [10]. Some of the reported carbon-based resistive memories rely on conductive bridge mechanism in which metal ions from the electrode diffuse into the insulating layer and form a conductive filament (CF) between the electrodes. This mechanism is indeed an electrochemical metallization (ECM) reaction and will not be covered in this work. Here, we specifically focus on carbon-based RRAM that relies on intrinsic carbon property switching. Carbon film comprises a mixture of $sp^2$ and $sp^3$ hybridization. The diamond-like $sp^3$ structure is high in resistance; whereas, the graphitic $sp^2$ structure is highly conductive. Pristine amorphous carbon film is $sp^3$ dominated and thus in high resistance state (HRS). After applying sufficiently high voltage ($V_{SET}$), distributed $sp^2$ nanochannels are formed across the disordered $sp^3$ carbon matrix as result of Joule heating effect [11,12]. These graphitic nanochannels provide low conductive paths and switch the device into low resistance state (LRS). Application of reset voltage ($V_{RESET}$) subsequently induces larger current and triggers the thermal fuse effect on these graphitic nanochannels [10]. This effectively ruptures the nanochannels and switches the device back into HRS. Figure 1 illustrated the evolution of the $sp^2$ nanochannels in HRS and LRS.

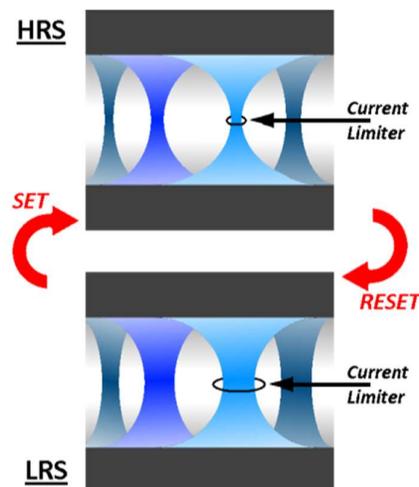

**Figure 1:** The schematic diagram illustrates the evolution of graphitic nanochannels when the RRAM device transition between low resistance state (LRS) and high resistance state (HRS).

The conductive nanochannels reassemble a hourglass structure [13] in which its conductivity is limited by the constriction. The constriction can be modelled as quantum wire (QW) and described within the framework of QPC model [14]. Despite the compactness of QPC model, parameter extraction through macroscopic I-V fitting is not as feasible compared with other typical conduction mechanisms in insulating films, *e.g.* Schottky emission, Poole-Frenkel emission, spare-charge-limited conduction (SCLC) or trap-assisted tunnelling (TAT) [15]. QPC model includes multiple nonlinear and closely coupled parameters, such as barrier height ($\Phi$), barrier curvature ($\alpha$) and voltage drop ratio ($\beta$). As a consequence, parameter extraction from this multi-dimensional model is formidable and potentially stuck at undesired local minimum [16]. Thus, in the work, we present a hybrid GA-based parameter extraction flow that able to extract multiple QPC parameters simultaneously based on experimental I-V characteristic.

## 2. Quantum Point Contact (QPC) Model

Schematic of a typical nanochannel is illustrated in Figure 2. The narrowest constriction of nanochannel limits current flow and determines the conductivity.

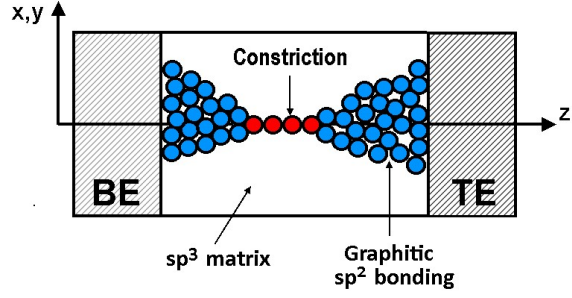

**Figure 2:** The schematic of a typical CF in unipolar RRAM.

The QPC model assumes the electrons transport over the constriction region as a quasi 1D system along the longitudinal direction. The dispersion curve of the electronic subbands is illustrated in Figure 3(a) and can be mathematically expressed as [17]:

$$E(k_z, z) = \epsilon_n(z) + \frac{\hbar^2 k_z^2}{2m} \quad (1)$$

where $z$ is the coordinates in the longitudinal direction, $k_z$ is the longitudinal coordinates in $k$-space, $m$ is electron mass, $\hbar$ is the reduced Planck constant and $\epsilon_n(z)$ can be expressed as:

$$\epsilon_n(z) = \frac{\pi^2 \hbar^2}{2m}\left(\frac{n_x^2}{L_x(z)^2} + \frac{n_y^2}{L_y(z)^2}\right) \quad (2)$$

The Equation 2 indicates that strong lateral confinement in the constriction of the CF could spaces out the subbands in the conduction band as shown in Figure 3(b). The bottom of subbands represents a potential barrier for the conduction path along $z$ direction. When lateral confinement at constriction area is strong, the bottom of the subbands is lifted and thus the potential barrier is raised. Figure 4 illustrates the potential barrier and the energy band diagram of the devices.

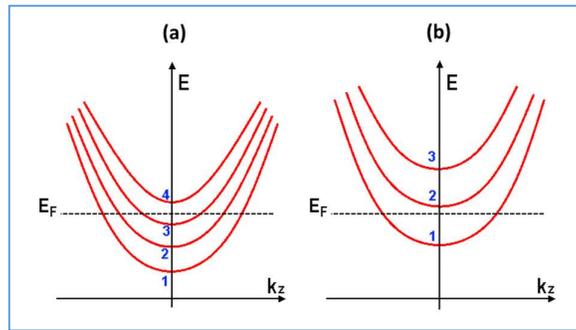

**Figure 3:** The dispersion curve of the electronic subbands (a) before lateral confinement (b) after lateral confinement.

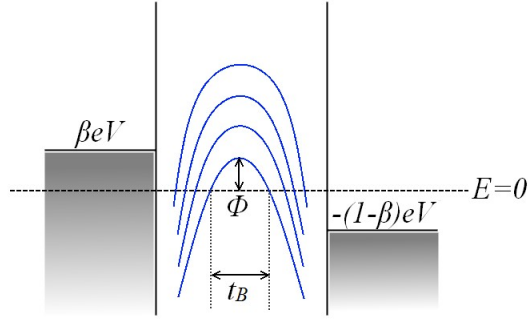

**Figure 4:** The potential barrier along the conduction path

Current flow across the narrow constriction of the CF can be calculated with Landauer transmission approach as shown in the Equation 3.

$$I = \frac{2e}{h} N_{ch} \times \int_{-\infty}^{\infty} T(E)[f(E - \beta eV) - f(E + (1-\beta)eV]dE \quad (3)$$

where $N_{ch}$ is the number of conducting channels, $e$ is electron charges, $h$ is Planck constant, $f$ is the Fermi-Dirac distribution and $T(E)$ is the transmission probability. The analytical expression for transmission probability can be expressed as in Equation 4.

$$T(E) = \frac{1}{1 + exp[-\alpha(E - \Phi)]} \quad (4)$$

In which $\Phi$ is the barrier height, $\alpha$ describes the curvature of the barrier and is related to barrier thickness ($t_B$) as shown in Equation 5.

$$\alpha = t_B \pi^2 \sqrt{\frac{2m^*}{\Phi}} \quad (5)$$

here, $t_B$ is the barrier thickness at the equilibrium Fermi energy. By inserting the transmission coefficient and expanding Equation 3, both HRS and LRS current can be represented in a compact form as in Equation 6.

$$I \approx \frac{2e}{h} N_{ch} \times \left\{ eV + \frac{1}{\alpha} ln\left[\frac{1 + exp\{\alpha[\Phi - \beta eV]\}}{1 + exp\{\alpha[\Phi + (1-\beta)eV]\}}\right] \right\} \quad (6)$$

here, the barrier height ($\Phi$) and barrier curvature ($\alpha$) describe the shape of the potential barrier and thus the resistance state of the memory device. Meanwhile, the fraction of voltage drop on cathode and anode is represented by $\beta$ and 1-$\beta$ respectively.

Unknown parameters in the QPC model are typically determined by assuming a fixed value of β and Φ. For instance, Lian *et al.* [18] extracted the QPC model parameters from HfO$_2$ based resistive memory by using fixed β and coupling parameter Φ with results of ab-initio simulations. β is set to 0.5 for symmetrical and β to 1 for asymmetrical voltage drop. This essentially reduced the equation to two fitting parameters, which are the number of vacancy paths (N) and average $t_{gap}$. However, voltage drop across the electrodes might not be perfectly symmetrical (β=0.5) or asymmetrical (β=1) in practice. As shown in Figure 5, 6 and 7, device conductance has a strong dependency with parameter β compared to other parameters, such as α and Φ. Therefore, prefixing the value of β in advance may undermine the accuracy of other extracted parameters. So, instead of extracting the QPC parameters directly by constraining β, a multi-variable constrained nonlinear search algorithm is required.

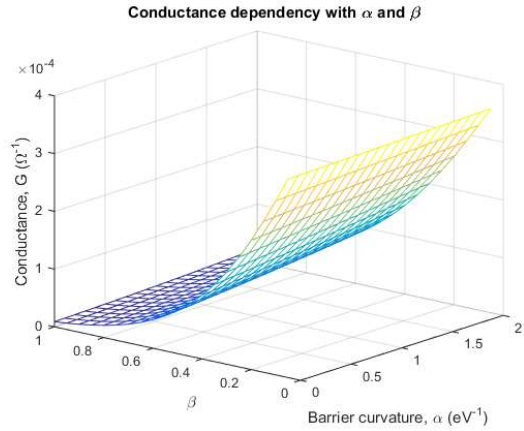

**Figure 5:** Conductance dependency with α and β.

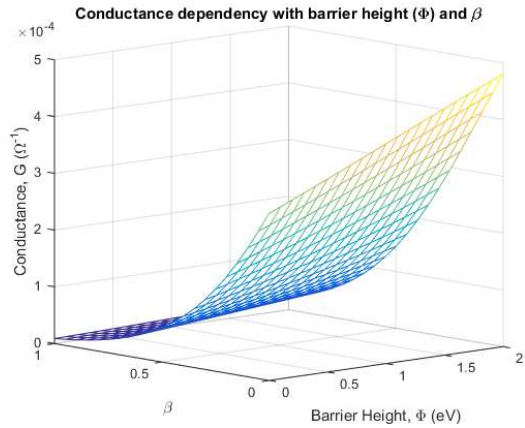

**Figure 6:** Conductance dependency with β and Φ.

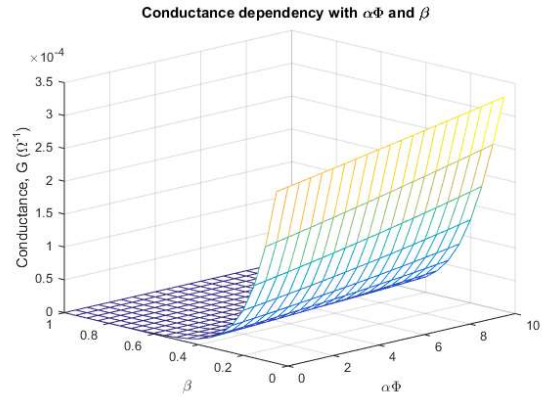

**Figure 7:** Conductance dependency with *β* and *αΦ*.

## 3. Hybrid Genetic Algorithm

Genetic algorithm (GA) is a directed search and optimization algorithm that resembles the natural evolution and biological processes of reproduction. GA can search and optimize multi-dimensional, non-differential and non-continuous problems. Common GA systems include four operators, i.e. (i) fitness function that describes the quality of chromosomes, (ii) selection operator that identifies chromosomes to reproduce, (iii) crossover operator that mates and breeds the next generation of chromosomes, and (v) mutation operator that introduces variation into offspring's chromosomes to get rid of the local minimum [19]. Figure 8 exhibits the flow of a typical GA execution.

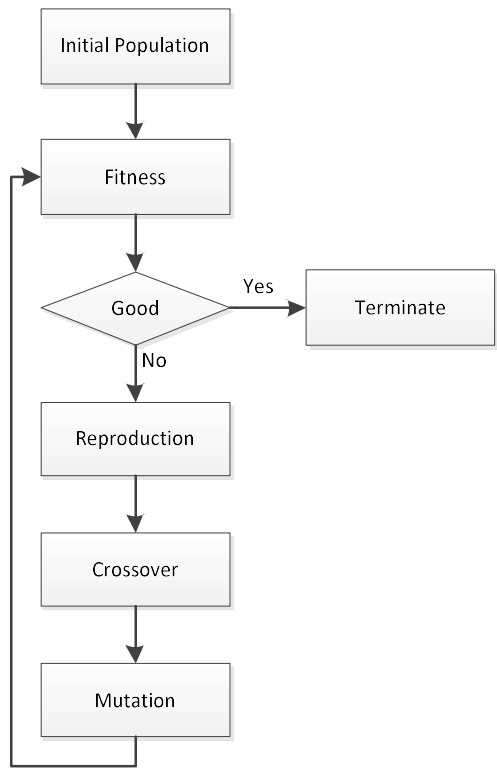

**Figure 8:** Typical flow for basic genetic algorithm

The GA flow begins by defining $N$ sets of chromosomes, where $N$ is often referred to as the population size. Each of these initial chromosomes is a random vector representing the parameters of the formulated problems, e.g. $i$-th chromosome can be expressed as:

$$\Omega_i = [p_1, p_2, p_3, \ldots, p_D] \tag{7}$$

where $p_x$ are the fitting parameters and $D$ represents the number of parameter that need to be extracted from the optimization problems. The fitness of the chromosome is then evaluated with a fitness function (Equation 8).

$$fit_i = f(\Omega_i) \tag{8}$$

Chromosomes with low fitness value will be discarded from the population. Meanwhile, the remaining good chromosomes will be paired to reproduce new offspring. There is a variety of selection methods. For example, the good chromosomes can be selected based on its normalized fitness. In this case, the probability of a particular chromosome to be selected to mate can be expressed with Equation 9.

$$P_i = \frac{fit_i}{\sum_{i=1}^{N} fit_i} \tag{9}$$

Once the mating candidates are identified, a crossover point is chosen randomly. For example, offspring 1 may inherit parameters at the left of the crossover point from parent 1, and the parameters at the right of crossover point from parent 2. Likewise, offspring 2 may inherit the remaining portion of the parameters from parent 1 and parent 2. The steps will be repeated until the stopping conditions are met.

GA is comparatively efficient in locating the vicinity of global optimum; however, it may take an abundant number of function evaluations to converge to the local optimum within the region. In this regard, hybridization of GA with a local search algorithm has been proved to be orders of magnitude better in terms of speed and accuracy [20]. By applying hybrid GA, the required generations of GA can be reduced substantially. The interim solution from the simple GA can be fine-tuned using another optimization solver that is more efficient in a local search, e.g. interior point or SQP algorithm.

## 4. Proposed QPC Parameter Extraction Flow

Some QPC Parameters such as the number of the single subband conduction channels in LRS ($N_{LRS}$) can be extracted directly from the boundary conditions. As the nanochannels are fully formed and the potential barrier is minimum in LRS, the current equation can be reduced to be linearly related to $V$. In this case, the conductance only depends on the total number of conduction channels and Equation 6 can be reduced to an Ohmic relation as in Equation 7.

$$I_{LRS} = \frac{2e^2}{h} \beta N_{LRS} V = G_0 \beta N_{LRS} V \tag{7}$$

where $G_0$ is the quantum conductance and $N_{LRS}$ represents the total number of single subband conduction channels in LRS. Hereby, we are able to express the $N_{LRS}$ in a function of paramater ($\beta$) and low biased experimental conductance in LRS.

$$N_{LRS} = \frac{1}{\beta G_0} \cdot \frac{I_{LRS(0)}}{V_{LRS(0)}} \tag{8}$$

where $V_{LRS(0)}$ is the smallest non-zero sampled voltage in the LRS experimental data and $I_{LRS(0)}$ is the corresponding current. For HRS, as the current is limited by a single conductive channel ($N_{HRS}=1$) with a parabolic barrier [14], the expression in Equation 6 can be reduced to Equation 9.

$$I_{HRS} = \frac{2e^2}{h}\exp(-\alpha_{HRS}\Phi)\left[V + \left(\frac{\alpha_{HRS}\beta}{2}\right)V^2\right] \tag{9}$$

Observe that the $I_{HRS}$ includes two components. The first term represents the Ohmic component with the equivalent transmission probability of $T(E) = exp(-\alpha\Phi)$. The second term is a nonlinear component dominating at high voltage. Therefore, if only low voltage regime is considered, the nonlinear component is negligible and the HRS current can be expressed as:

$$I_{HRS(0)} = G_0 N_{HRS} \exp(-\alpha_{HRS}\Phi) V_{HRS(0)} \tag{10}$$

where $V_{HRS(0)}$ is the smallest sample voltage in the HRS experimental data and $I_{HRS(0)}$ is the corresponding current. By replacing $N_{HRS}$ to one, the barrier curvature parameter at HRS ($\alpha_{HRS}$) can be represented with a function of barrier height ($\Phi_{HRS}$).

$$\alpha_{HRS} = -\frac{1}{\Phi_{HRS}} \ln\left[\frac{I_{HRS(0)}}{V_{HRS(0)}} \cdot \frac{1}{G_0}\right] \tag{11}$$

With these proposed steps, the formula could be reduced to only four unknown parameters, i.e. $\Phi_{HRS}$, $\Phi_{LRS}$, $\alpha_{LRS}$ and $\beta$. Both LRS and HRS data can then be fitted simultaneously using simple GA. The fitness function is defined as the average of mean absolute percentage error (MAPE) for both HRS and LRS data:

$$\begin{aligned}fitness &= 0.5 * MAPE_{LRS} + 0.5 * MAPE_{HRS} \\ &= \frac{1}{2M}\sum_{i=1}^{M}\left|\frac{I_{LRS(i)} - I_{QPC-LRS(i)}}{I_{LRS(i)}}\right| \\ &+ \frac{1}{2N}\sum_{i=1}^{N}\left|\frac{I_{HRS(i)} - I_{QPC-H\ (i)}}{I_{HRS(i)}}\right|\end{aligned} \tag{12}$$

where $I_{LRS}$ and $I_{HRS}$ are experimental currents at LRS and HRS respectively. $I_{QPC-LRS}$ and $I_{QPC-HRS}$ are the simulated current under the corresponding voltage. The simple GA is repeated until the relaxed stopping criteria are met. The interim solution is then to be set as the initial solution for the subsequent interior point local search algorithm. The proposed QPC model parameter extraction flow is summarized in Figure 9.

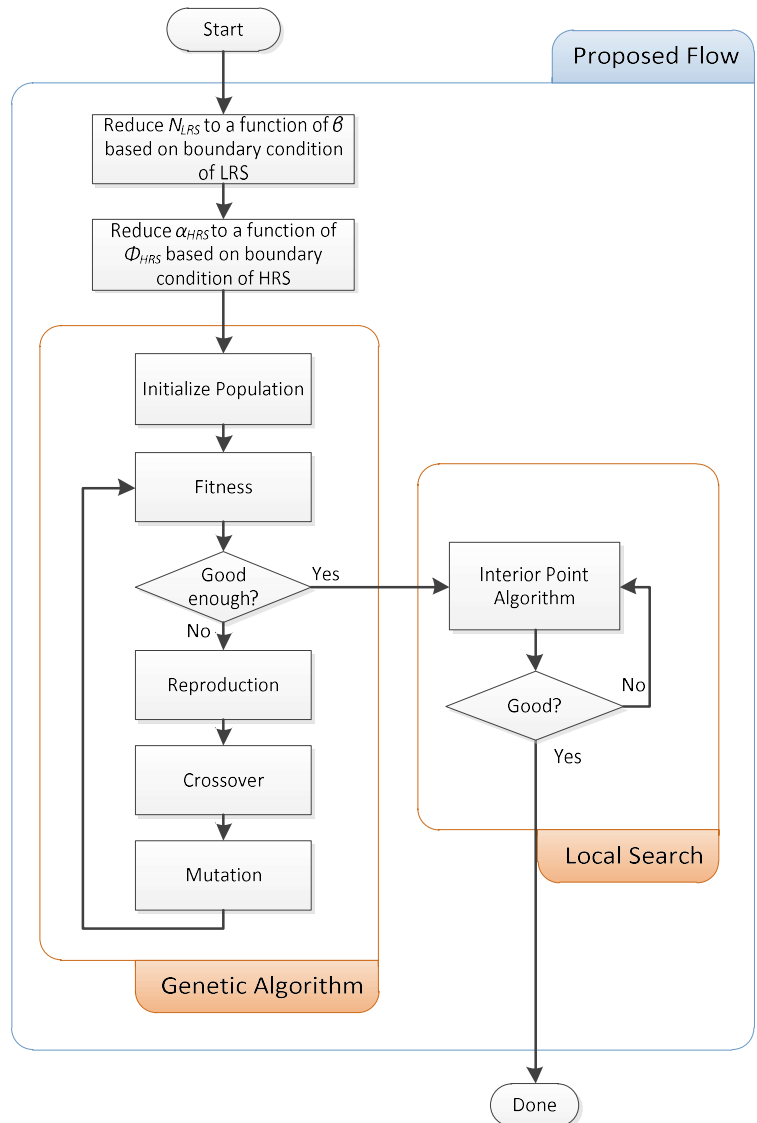

**Figure 9:** The proposed QPC model extraction flow.

The proposed method is applied to extract QPC model parameters of a diamond-like carbon (DLC)-based resistive memory [21]. The initial population size and the fitness function tolerance for GA optimization are set to 50 and 0.001% respectively. The GA-only flow takes 283 generations to achieve the MAPE of 3.72%. On the other hand, the hybrid GA flow successfully reaches the equivalent accuracy with only 25 generations and 43 iterations of the interior point algorithm. The convergence profile of both GA-only and hybrid GA flows are depicted in Figure 10.

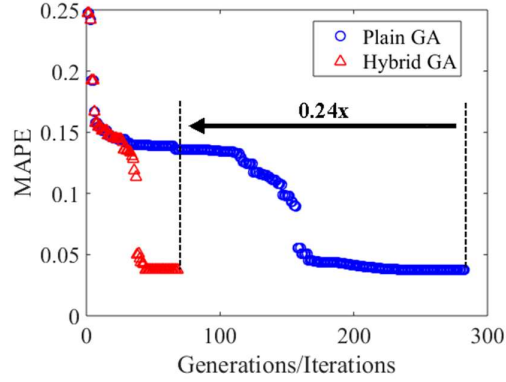

**Figure 10:** Convergence profile of the GA-only (blue) and the hybrid GA flow with interior-point local search algorithm (red). Note that the plain GA flow takes 283 generations to reach the MAPE of 3.7%, while hybrid GA flow takes only 25 generations of GA and 43 iterations of interior point local search.

The hybrid GA flow locates the interim solution with MAPE of 14.4% in the 25th generations and the interior point algorithm is then applied to converge the interim solution to its local minimum. Note that each generation of GA incurs 50 function calls (equals to population size), while the interior point algorithm takes average 7 function calls each iterations. Therefore, by applying this proposed hybrid GA flow, the QPC Parameters can be extracted with only fraction of functions calls.

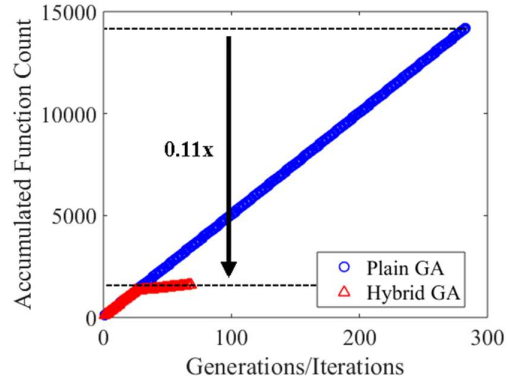

**Figure 11:** The GA-only flow made accumulated function calls of 14200; meanwhile, the hybrid GA flow takes merely 1613 function calls to achieve the same accuracy.

The extracted QPC parameters are listed in Table 1. As shown in Figure 12, the resulting QPC model fits excellently with the experimental data. Note that the extracted $\beta$ is not a perfect "1", this substantiates the assumption of perfect asymmetric voltage drop ($\beta$=1) may undermine the accuracy of other extracted parameters. The extracted barrier height in HRS ($\Phi_{HRS}$) is within the 1-4 eV range commonly observed in DLC film [22].

**Table 1:** Extracted parameters for QPC model.

| Reference | Devices | Extracted QPC Parameters | | | | | MAPE |
|---|---|---|---|---|---|---|---|
| | | $\Phi_{HRS}$ | $\Phi_{LRS}$ | $\alpha_{HRS}$ | $\alpha_{LRS}$ | $\beta$ | |
| J. Xu et al. [21] | Pt/DLC/Pt | 2.9276 eV | 0.056 eV | 1.988 eV$^{-1}$ | 242.44 eV$^{-1}$ | 0.9693 | 3.72% |

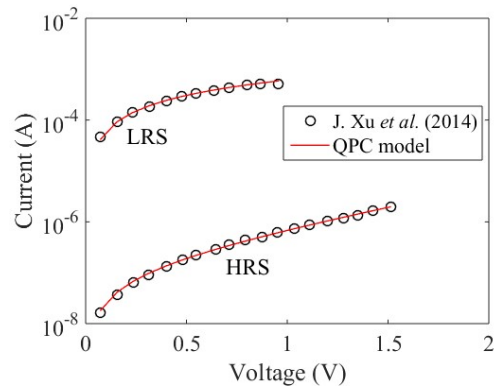

**Figure 12:** QPC model fitting with experimental data [21].

## 5. Conclusion

In this paper, the conduction mechanism of carbon-based resistive memories is explained within the framework of QPC model. Landauer theory for the mesoscopic system is applied to model current transport in the graphitic $sp^2$ nanochannels. The resistance state of the memory device is correlated to the density of nanochannels. In this work, an efficient hybrid GA based QPC model parameter extraction flow is proposed. The QPC model parameters are extracted from experimental data by using the GA together with interior point algorithm. The proposed hybrid flow reduces the accumulated function calls from 14200 to 1613 by introducing an efficient local search algorithm. The simple GA flow is used to locate the proximity of the global minimum while the interior point algorithm is used to refine and locate the optimum solution. Our results indicate that the hybrid GA is faster and more efficient in parameter extraction compared to the homogeneous GA. The extracted parameters are found to be in appropriate range and the MAPE of the resulting QPC models is only 3.72% compared to the experimental results. Hereby, this work affirms that the QPC model is a feasible model to describe the conduction mechanism of graphitic nanochannels in DLC-based resistive memories.

## References

1. Wang, L.; Yang, C.-H.; Wen, J. Physical principles and current status of emerging non-volatile solid state memories. *Electron. Mater. Lett.* **2015**, *11*, 505–543.

2. Meena, J. S.; Sze, S. M.; Chand, U.; Tseng, T.-Y. Overview of emerging nonvolatile memory technologies. *Nanoscale Res. Lett.* **2014**, *9*, 526.

3. Chua, L. Resistance switching memories are memristors. *Appl. Phys. A Mater. Sci. Process.* **2011**, *102*, 765–783.

4. Kreupl, F. Carbon Memory Assessment. *arXiv Prepr. arXiv1408.4600* **2014**, 1–15.

5. Zhuge, F.; Hu, B.; He, C.; Zhou, X.; Liu, Z.; Li, R. W. Mechanism of nonvolatile resistive switching in graphene oxide thin films. *Carbon N. Y.* **2011**, *49*, 3796–3802.

6. Wang, L. H.; Yang, W.; Sun, Q. Q.; Zhou, P.; Lu, H. L.; Ding, S. J.; Wei Zhang, D. The mechanism of the asymmetric SET and RESET speed of graphene oxide based flexible resistive switching memories. *Appl. Phys. Lett.* **2012**, *100*, 3–7.

7. Zhuge, F.; Dai, W.; He, C. L.; Wang, a. Y.; Liu, Y. W.; Li, M.; Wu, Y. H.; Cui, P.; Li, R. W. Nonvolatile resistive switching memory based on amorphous carbon. *Appl. Phys. Lett.* **2010**, *96*, 2–4.

8. Chen, Y.-J.; Chen, H.-L.; Young, T.-F.; Chang, T.-C.; Tsai, T.-M.; Chang, K.-C.; Zhang, R.; Chen, K.-H.; Lou, J.-C.; Chu, T.-J.; Chen, J.-H.; Bao, D.-H.; Sze, S. M. Hydrogen induced redox mechanism in amorphous carbon resistive random access memory. *Nanoscale Res. Lett.* **2014**, *9*, 52.

9. Dellmann, L.; Sebastian, A.; Jonnalagadda, P.; Santini, C. a.; Koelmans, W. W.; Rossel, C.; Eleftheriou, E. Nonvolatile resistive memory devices based on hydrogenated amorphous carbon. *2013 Proc. Eur. Solid-State Device Res. Conf.* **2013**, *1*, 268–271.

10. Fu, D.; Xie, D.; Feng, T. T.; Zhang, C. H.; Niu, J. B.; Qian, H.; Liu, L. T. Unipolar Resistive Switching Properties of Diamondlike Carbon-Based RRAM Devices. *IEEE Electron Device Lett.* **2011**, *32*, 803–805.

11. Evtukh, A.; Litovchenko, V.; Semenenko, M.; Yilmazoglu, O.; Mutamba, K.; Hartnagel, H. L.; Pavlidis, D. Formation of conducting nanochannels in diamond-like carbon films. *Semicond. Sci. Technol.* **2006**, *21*, 1326–1330.

12. Lim, E. W.; Ahmadi, M. T.; Ismail, R. Modeling and simulation of graphene-oxide-based RRAM. *J. Comput. Electron.* **2016**, *15*, 602–610.

13. Lee, B. H.; Bae, H.; Seong, H.; Lee, D. Il; Park, H.; Choi, Y. J.; Im, S. G.; Kim, S. O.; Choi, Y. K. Direct Observation of a Carbon Filament in Water-Resistant Organic Memory. *ACS Nano* **2015**, *9*, 7306–7313.

14. Miranda, E. A.; Walczyk, C.; Wenger, C.; Schroeder, T. Model for the Resistive Switching Effect in HfO2 MIM Structures Based on the Transmission Properties of Narrow Constrictions. *IEEE Electron Device Lett.* **2010**, *31*, 609–611.

15. Lim, E. W.; Ismail, R. Conduction Mechanism of Valence Change Resistive Switching Memory: A Survey. *Electronics* **2015**, *4*, 586–613.

16. International Technology Roadmap for Semiconductors (ITRS) *International Technology Roadmap for Semiconductors - Modeling and Simulation*; 2013.

17. Brandbyge, M.; Schiotz, J.; Sorensen, M. R. Atom-Sized Wires Between Two Metals. *Phys. Rev. B* **1995**, *52*.

18. Lian, X.; Cartoixà, X.; Miranda, E.; Perniola, L.; Rurali, R.; Long, S.; Liu, M.; Suñé, J. Multi-scale quantum point contact model for filamentary conduction in resistive random access memories devices. *J. Appl. Phys.* **2014**, *115*, 244507.

19. Carr, J. An Introduction to Genetic Algorithms. **2014**, 1–40.

20. Grosan, C.; Abraham, A.; Ishibuchi, H. Hybrid Evolutionary Algorithms: Methodologies, Architectures, and Reviews. In *Hybrid Evolutionary Algorithms*; Springer Berlin Heidelberg, 2007; pp. 1–17.

21. Xu, J.; Xie, D.; Peng, P.; Zhang, X.; Ren, T. Bipolar Resistive Switching Behaviros in An Al/DCL/W Structure. In *ICSICT2014*; 2014; pp. 1–3.

22. Robertson, J. Gap states in diamond-like amorphous carbon. *Philos. Mag.* **1997**, *76*, 335–350.